%% file: paper.tex
\documentclass[useAMS,usenatbib]{mn2e}

\usepackage{graphicx}
\usepackage{aas_macros}
\usepackage{amssymb,amsmath}

\title{Stacking catalog sources in WMAP data}

\author[K.~W.~Schultz and K.~M.~Huffenberger]{K.~W.~Schultz and K.~M.~Huffenberger\thanks{Email: huffenbe@physics.miami.edu}\\
Department of Physics, University of Miami, 1320 Campo Sano Dr., Coral Gables, FL 33146, USA}

\newlength{\figwidth}
\begin{document}
\maketitle

\label{firstpage}

\begin{abstract}
We stack WMAP 7-year temperature data around extragalactic point sources, showing that the profiles are consistent with WMAP's beam models, in disagreement with the findings of \citet{2010MNRAS.tmpL..93S}.  These results require that the source sample's selection is not biased by CMB fluctuations.  We compare profiles from sources in the standard WMAP catalog, the WMAP catalog selected from a CMB-free combination of data, and the NVSS catalog, and quantify the agreement with fits to simple parametric beam models.  We estimate the biases in source profiles due to alignments with positive CMB fluctuations, finding them roughly consistent with those biases found with the WMAP standard catalog.  Addressing those biases, we find source spectral indices significantly steeper than those used by WMAP, with strong evidence for spectral steepening above 61 GHz.  Such changes modify the power spectrum correction required for unresolved point sources, and tend to weaken somewhat the evidence for deviation from a Harrison-Zel'dovich primordial spectrum, but more analysis is required.  Finally, we discuss implications for current CMB experiments.

\end{abstract}


\section{Introduction}

A telescope's beam or point spread function, which dampens the instrumental response to fluctuations on small scales, is prominent among the systematic effects that must be well-understood for Cosmic Microwave Background (CMB) measurements.
  The act of observing convolves the sky with the telescope beam, so observations of bright objects that are point-like (compared to the beam size) provide an obvious check on the beam pattern.  The Wilkinson Microwave Anisotropy Probe (WMAP) team used observations of Jupiter to measure the beams \citep{2011ApJS..192...14J,2009ApJS..180..246H,2007ApJS..170..263J,2003ApJS..148...39P}, but in principle any suitable objects will work, including extragalactic point sources.

\citet{2010MNRAS.tmpL..93S} used the  5-year WMAP data to construct stacked profiles around sources detected in the WMAP point source catalog \citep{2009ApJS..180..283W}.  \cite{2011arXiv1107.2654W} repeated this analysis with the 7-year data, using  WMAP and external catalogs.  Both papers report some intriguing discrepancies: for WMAP's differencing assemblies at 40--90 GHz, they found substantial offsets at large scales, and the profiles appear broader than the beam patterns from the Jupiter measurements.  They address and discard several possible explanations for the effect: extended radio sources, source clustering, and selection bias near the catalog threshold, and favor a nonlinearity in WMAP's response to Jupiter, which has a peak temperature $\sim 3$ orders of magnitude higher than CMB fluctuations at these frequencies.  They note that a cosmological analysis based on the window functions computed from their stacked profiles (instead of the Jupiter model) would significantly change WMAP's basic cosmological results, for example by changing the height and location of the first acoustic peak in the power spectrum.  In \citet{2010A&G....51e..14S}, the authors present this finding as a challenge to the $\Lambda$CDM paradigm.

Despite these findings, several contrary lines of evidence indicate that the Jupiter-based beam models sufficiently represent the true beam patterns for the WMAP telescope.   First, WMAP and several pre- and post-WMAP experiments (from the ground and from balloons) have over the past decade found similar CMB power spectra on the scales where they overlap.  Telescopes for which the beam scale differs substantially from WMAP are particularly powerful probes of this consistency (see WMAP comparisons to
QUaD: \citealt{2009ApJ...705..978B};
ACBAR: \citealt{2009ApJ...694.1200R};
ACT: \citealt{2010arXiv1009.0777H};
and SPT: \citealt{2011arXiv1105.3182K}).  
Second, using WMAP data alone, comparison of the power spectra derived from the different differencing assemblies can be a useful test of errors in the beam models.  A frequency-dependent beam systematic error of the size considered by \citet{2010MNRAS.tmpL..93S} would almost certainly show up in estimates for the unresolved point source contribution at high-$l$.  \citet{2009ApJS..180..296N} and \citet{2008ApJ...688....1H} extensively studied multifrequency combinations of Jupiter-beam-corrected power spectra from WMAP for this purpose, and found no such substantial beam anomaly.  These methods constrain the relative beam window functions between different assemblies at roughly the percent level. For example, a bump at $l<200$ in the WMAP 3-year data's unresolved point source estimate disappeared with the beam revision of \citet{2009ApJS..180..246H} for the 5-year data.  This revision changed the measurement of first acoustic peak's height by 2--3 percent, much smaller than the change in the beam proposed by \citet{2010MNRAS.tmpL..93S} or \cite{2011arXiv1107.2654W}.   (See also \citealt{2010A&G....51e..16G} for further  discussion of the strengths of the $\Lambda$CDM model despite the misgivings of \citealt{2010A&G....51e..14S}.)

However, the issue with the stacked point sources remains: why should stacked extragalactic point sources present a profile so different from the telescope beam?  This is a worthwhile question.  Here, we explore it by stacking sources in the same way as \citet{2010MNRAS.tmpL..93S}, and we are able to reproduce their basic results for the WMAP standard catalog.  However, by using alternative source catalogs, we by contrast find source profiles compatible with the WMAP beam models.  We fit parameterized models to quantify the beam effects, and further explore systematic selection biases.  In section \ref{sec:methods}, we present our data selection and analysis methods, while in section \ref{sec:results} we discuss and interpret our results.  Finally in section \ref{sec:conclusions} we draw our conclusions.

\section{Methods} \label{sec:methods}

\subsection{Data selection}
We base our analysis on the WMAP 7-year maps and point source catalogs, available from the LAMBDA website\footnote{\tt http://lambda.gsfc.nasa.gov/}.  In turn we use maps for the individual differencing assemblies (DAs), with and without a foreground template removed.  We focus on Q-band (41 GHz), V-band (61 GHz), and W-band (94 GHz).  All maps are at HEALPix\footnote{\tt http://healpix.jpl.nasa.gov} resolution $N_{\rm side} = 512$ ($6.9'$ pixels). 

We mask the sky to exclude pixels and sources near the galactic plane or other extended structures (like the LMC), while retaining bright sources away from extended foregrounds.  We begin with the WMAP 7-yr temperature analysis mask.  Then we invert the WMAP point source mask to retain those pixels near sources excluded by temperature analysis mask.  The negative side effect is that some pixels are now kept around sources in high-foreground regions.  To eliminate these, we smooth with a $2^{\circ}$ FWHM Gaussian and apply a threshold at 90 percent of the smoothed maximum.  This expands the mask slightly, eliminating the problem pixels.  Our resultant mask excludes the highest foreground regions and leaves 76.8 percent of the sky available for our analysis.

The standard WMAP catalog contains 471 sources, but our mask excludes 38.  Closely paired sources can spoil the profile, so we additionally require that the sources be isolated.  Since we explore the source profiles out to $\theta_{\rm max} = 1.0^{\circ}$, we exclude all catalog sources which are separated by less than $2\theta_{\rm max}$ from another source.  (Both members of the pair are excluded.)  This prevents overlaps in computing our stacked profiles, and removes an additional 68 sources from our analysis (20 of these sources have another source even within $1.0^{\circ}$).  After the two cuts, we have 365 sources on which we base our stacking analysis.  The positional uncertainty in the WMAP catalog is $4'$ \citep{2003ApJS..148...97B,2008ApJ...681..747C}.

The WMAP team also provides a catalog selected from CMB-subtracted maps using the multifrequency method of \citet{2008ApJ...681..747C}. The catalog contains 417 sources. Our mask eliminates 52 sources, then we cull 34 members of close pairs, leaving 331.  The positional uncertainty in the CMB-free catalog is smaller, about $2'$.  Of these 331 sources, 260 (or 78.5\%) lie within $0.3^{\circ}$ of a WMAP standard catalog source.

Finally, we check our results by stacking on the source catalog from the NRAO VLA Sky Survey \citep[NVSS,][]{1998AJ....115.1693C}, which surveyed the sky at 1.4 GHz for  $\delta > -40^\circ$.  At that frequency, the CMB is not a significant component of the emission, and cannot affect the catalog selection.  We cut at 2 Jy (at 1.4 GHz) to take the brightest 762 sources.  Masking eliminates 350 sources, and culling close pairs eliminates another 129 sources, leaving 283.  Accounting for the reduced sky coverage, this gives a similar density of sources as the WMAP catalogs.  The positional uncertainty of the bright NVSS sources is very small, $< 1''$.  Of these 283 NVSS sources, a total of 71 (or 25.1\%) lie within $0.2^{\circ}$ of a WMAP standard catalog source.  Thus NVSS sources represent a significantly distinct population.

\subsection{Stacked profiles}

We stack our sources simply by defining angular annuli around reported catalog positions, then averaging the pixels whose centers fall into those annuli. 

The shape of the profile is subject to pixelization effects, both from the angular binning into annuli and from the map pixelization.  WMAP's beam models are tabulated at much higher angular resolution than the profiles, so must also be binned for direct comparison to the profile, or for the computation of $\chi^2$.  
\begin{figure}
\begin{flushright}
\includegraphics[width=0.97\columnwidth]{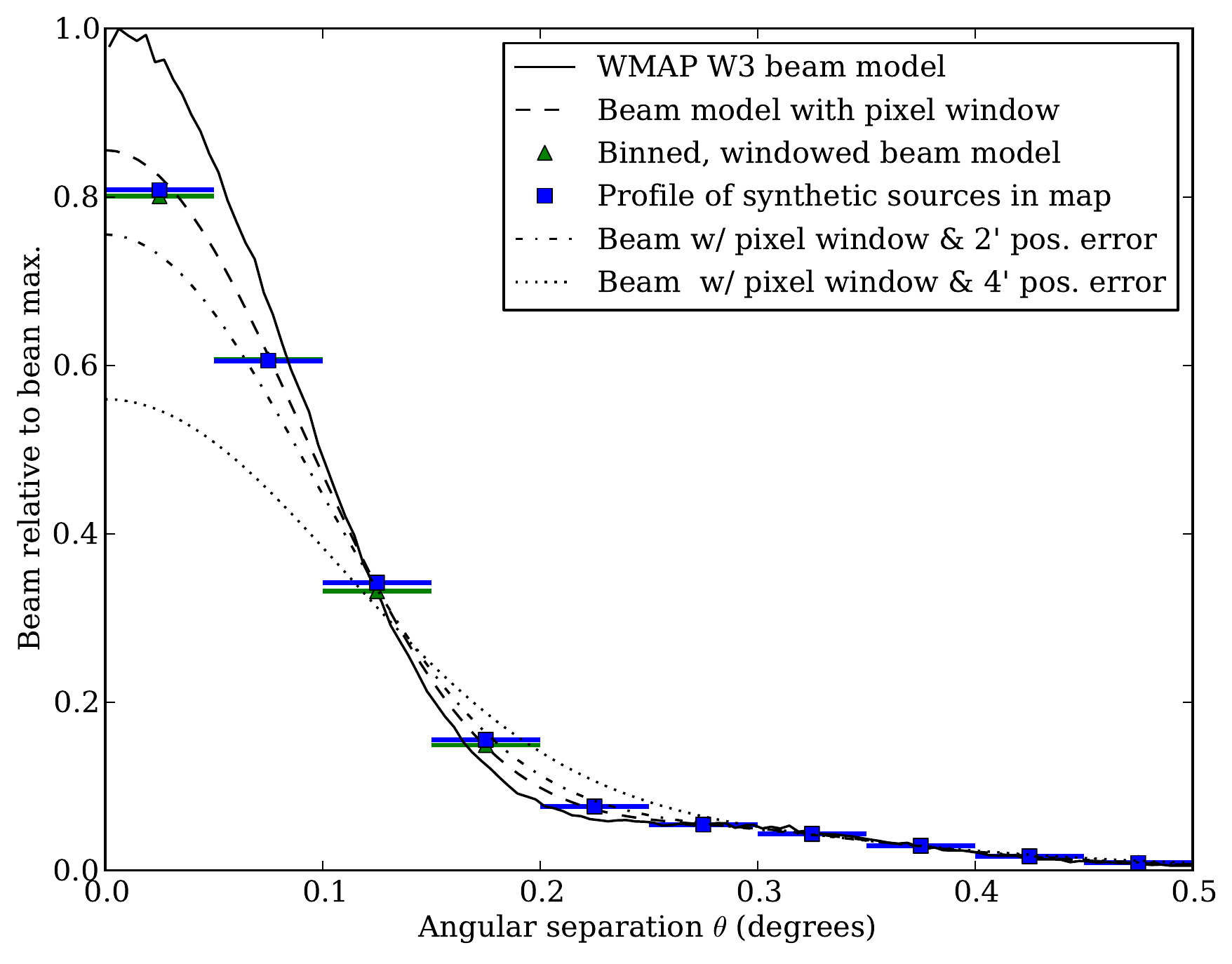}
\end{flushright}
\caption{Smearing and pixelization effects compared to the input beam model for the W3 DA.  After convolution with the HEALPix pixel window function, we show the binning of the input beam into annuli, where horizontal bars indicate the bin width.  This is very close to the stacked profile from a Monte Carlo of synthetic sources.  We also plot the beam model convolved with a $\sigma = 2'$ or $4'$ Gaussian to represent uncertainty in the source position.}\label{fig:pixel_effects}
\end{figure}

On an infinite resolution map, binning will simply average the beam model ($b$) over annuli (indexed by $i$) to get the binned model ($m$):
\begin{equation}
  m_i = { \int_{\theta_{{\rm min},i}}^{\theta_{{\rm max},i}} d\theta\ \theta\ b(\theta) } \left/ { \int_{\theta_{{\rm min},i}}^{\theta_{{\rm max},i}} d\theta\ \theta}. \right. 
\end{equation}
The angular weighting causes the peak to be slightly suppressed.  

For point sources in realistic maps, the map pixel size and shape also influence the profile, broadening and blunting the peak.  This affects the smallest beams the most.  To simulate this effect, we placed 1000 synthetic sources with the shape of each WMAP beam at random locations in a high-resolution HEALPix map, then reduced the resolution to $N_{\rm side} = 512$ and computed a stacked profile around the source center.  (We used $N_{\rm side} = 4096$ for the high-resolution map; using $N_{\rm side} = 2048$ makes only a minor difference at W band, and is immaterial for V and larger beams.)  This results in a profile which is notably blunted at the peak, and then larger than the input beam at $0.2^\circ \lesssim \theta \lesssim 0.4^\circ$.   At the peak, pixel effects suppress the  W band beams slightly more than ten percent, V band by about five percent, Q by three percent, and the K and Ka band profiles by less than two percent.  Normalizing the pixelized, blunted profiles to the unbinned beam model at the peak can make the binned profile's tail look too heavy, but this alone is insufficient to account for the broad profiles seen by \citet{2010MNRAS.tmpL..93S}.
The source profiles in maps are well-reproduced by convolving the beam with the map pixel window function and then binning.  This is the strategy we use later for our parametric beam models.  We illustrate these binning and pixel effects in Figure~\ref{fig:pixel_effects} for the W3 DA.  The W-band beams are WMAP's narrowest in the main lobe and this particular beam has an interesting shape, with a shoulder from $0.2$--$0.4$ degrees.  We model the uncertainty in source position by convolving the profile with a Gaussian of appropriate size ($\sigma = 4'$ for the standard WMAP catalog, $\sigma = 2'$ for the CMB-free catalog, and $\sigma=0$ for the NVSS catalog).

We consider two contributions to the covariance matrix for these profiles, from detector noise and from background CMB fluctuations.  Our final covariance matrix is the sum of these components.  Below we use these covariance matrices when minimizing $\chi^2$ for model fitting.  We not make any correction for the finite number of sources stacked in the profile, which will slightly modify the errors because of positional uncertainty (deviating from the Gaussian convolution kernel appropriate for an infinite number of sources).

For non-overlapping source profiles, under the assumption of white noise, the detectors contribute a diagonal component to the covariance.  This term we compute analytically from the noise variance per pixel provided by the WMAP team for each differencing assembly's map (Appendix~\ref{sec:covariance_app}).

Under the assumption that source positions are uncorrelated with CMB fluctuations, we can construct the covariance due to the background CMB.  For the WMAP standard catalog, we will later see that this assumption is unsound.  The covariance can be computed analytically, but is more practically computed with a Monte Carlo method, as follows.

For the CMB power spectrum, we use the WMAP best-fit $\Lambda$CDM model.  We compute stacked profiles on CMB-only simulations (including the beam and pixel window functions) and combine them to produce our estimate.  We use 1600 Monte Carlo realizations, although using half that number changes our typical estimate of $\chi^2$ by just one half of one percent.

\begin{figure}
\begin{center}
\includegraphics[width=\columnwidth]{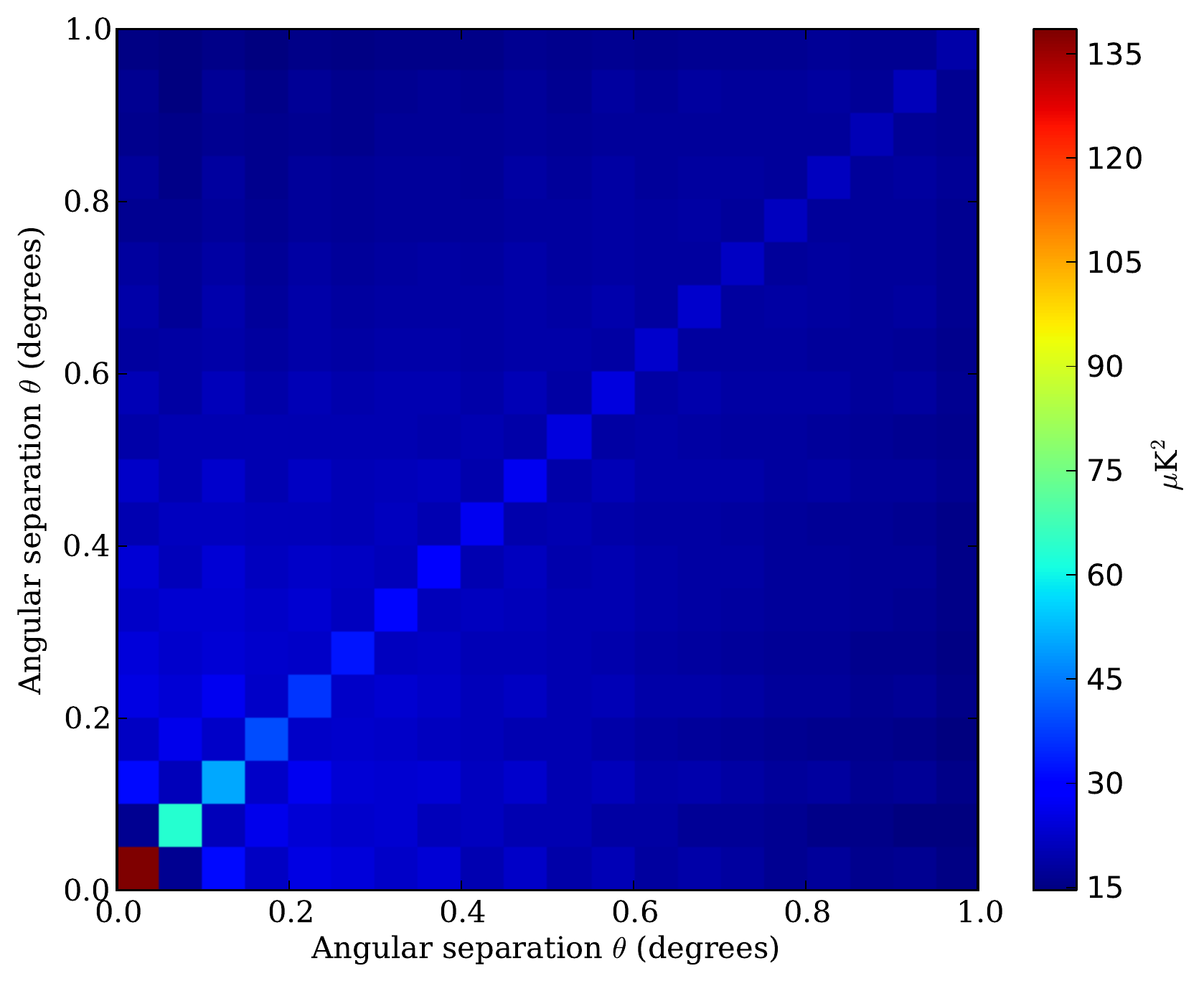}
\end{center}
\caption{The covariance matrix due to CMB and noise fluctuations for a profile in the W3 differencing assembly, using the standard WMAP catalog.  Other DAs are qualitatively similar. CMB fluctuations correlate across the range of angular separations, while white noise contributes only to the diagonal, and is suppressed by the larger number of observations in annuli further from the center.}\label{fig:profile_covariance}
\end{figure}

Unlike the noise, the covariance due to CMB fluctuations is strongly correlated between profile bins (Figure~\ref{fig:profile_covariance}).

The covariance matrices for the standard WMAP catalog and the WMAP CMB-free catalog are very similar because they share a large number of sources.  The covariance for NVSS sources is somewhat larger for two main reasons.  First, the number of NVSS sources is smaller.  Second, NVSS did not survey the South ecliptic pole ($\delta \approx -66.6$).  WMAP noise is notably suppressed at the ecliptic poles due to a higher number of observations, so the NVSS profiles are taken from higher noise portions of the sky.  This increases the errors accordingly below.

\subsection{Minimizing $\chi^2$ for Model Fitting}
Below we explore parametric models for the profile, typically applying an amplitude factor and an offset.  To fit a model to our stacked source profile we minimize:
\begin{equation}
\chi^{2}(\alpha)  =  \sum_{ij} [ P_i - m_i(\alpha) ]  C^{-1}_{ij} [ P_j - m_j(\alpha)].
\end{equation}
Here $P$ is the stacked profile and $m$ is a binned model for the profile, which in turn is based on the WMAP beam, the HEALPix pixel window function, and the positional uncertainty appropriate for the catalog.  The indices $i,j$ run over the angular bins.  This model additionally depends on a set of parameters $\alpha$ (like an amplitude, etc.).  
When the model depends on the parameters linearly, we can solve for the best fit parameters and their covariance matrix algebraically.  Otherwise we use an implementation of Powell's direction sets method to minimize $\chi^2$ for nonlinear models.  We compute parameter covariances with
\begin{equation}
  [{\rm Cov}(\alpha_p,\alpha_q)]^{-1} =   \sum_{ij} \frac{\partial m_i}{\partial \alpha_p} C^{-1}_{ij} \frac{\partial m_j}{\partial \alpha_q},
\end{equation}
although this is strictly only applicable when the fit is good.

\section{Results} \label{sec:results}

\subsection{Stacked source profiles}
\begin{figure*}
\includegraphics[width=\columnwidth]{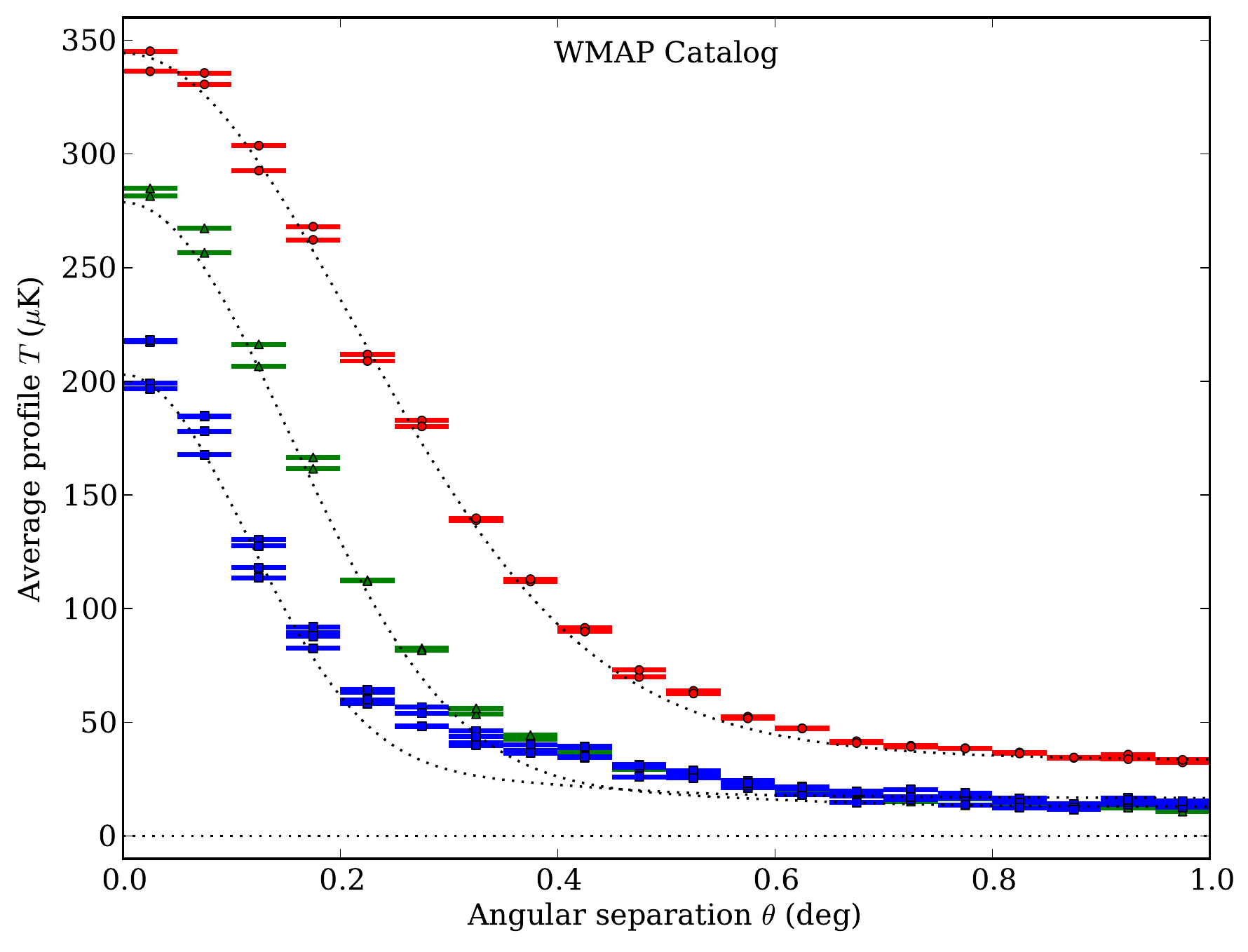}
\hfill
\includegraphics[width=\columnwidth]{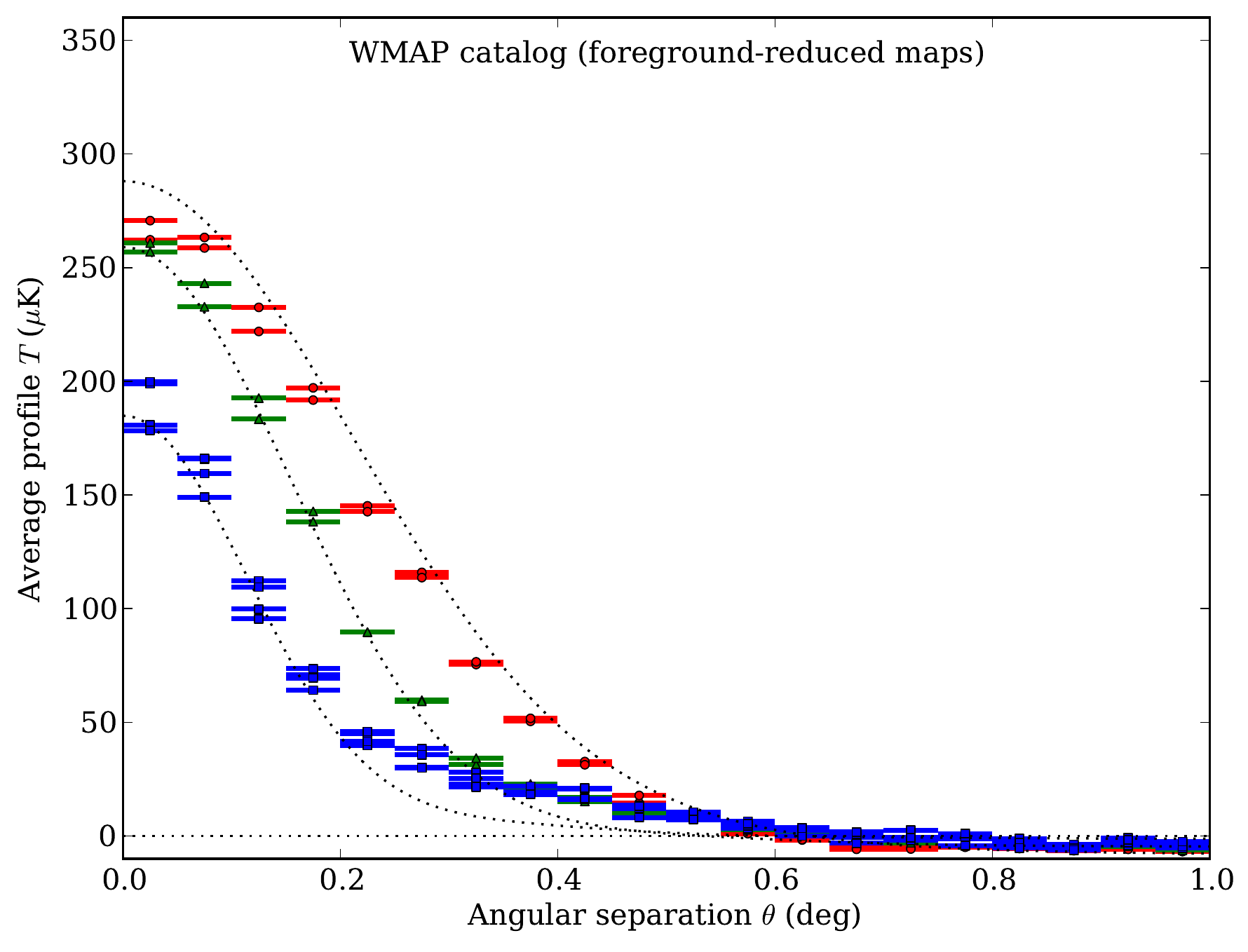} 
\caption{Stacked profiles around sources from the WMAP point source catalog for all DAs, where the horizontal bars show the angular bin width.  The source amplitude decreases from Q-band (red circles) to V-band (green triangles) to W-band (blue squares).  The best-fit models for Q1, V1, and W1 are shown as dotted lines, and are poor fits to the profiles.  Left:  using the raw WMAP  maps for each differencing assembly.  Right:  the same, but using foreground-reduced maps.} \label{fig:stacked_profiles}
\end{figure*}
\begin{figure*}
\includegraphics[width=\columnwidth]{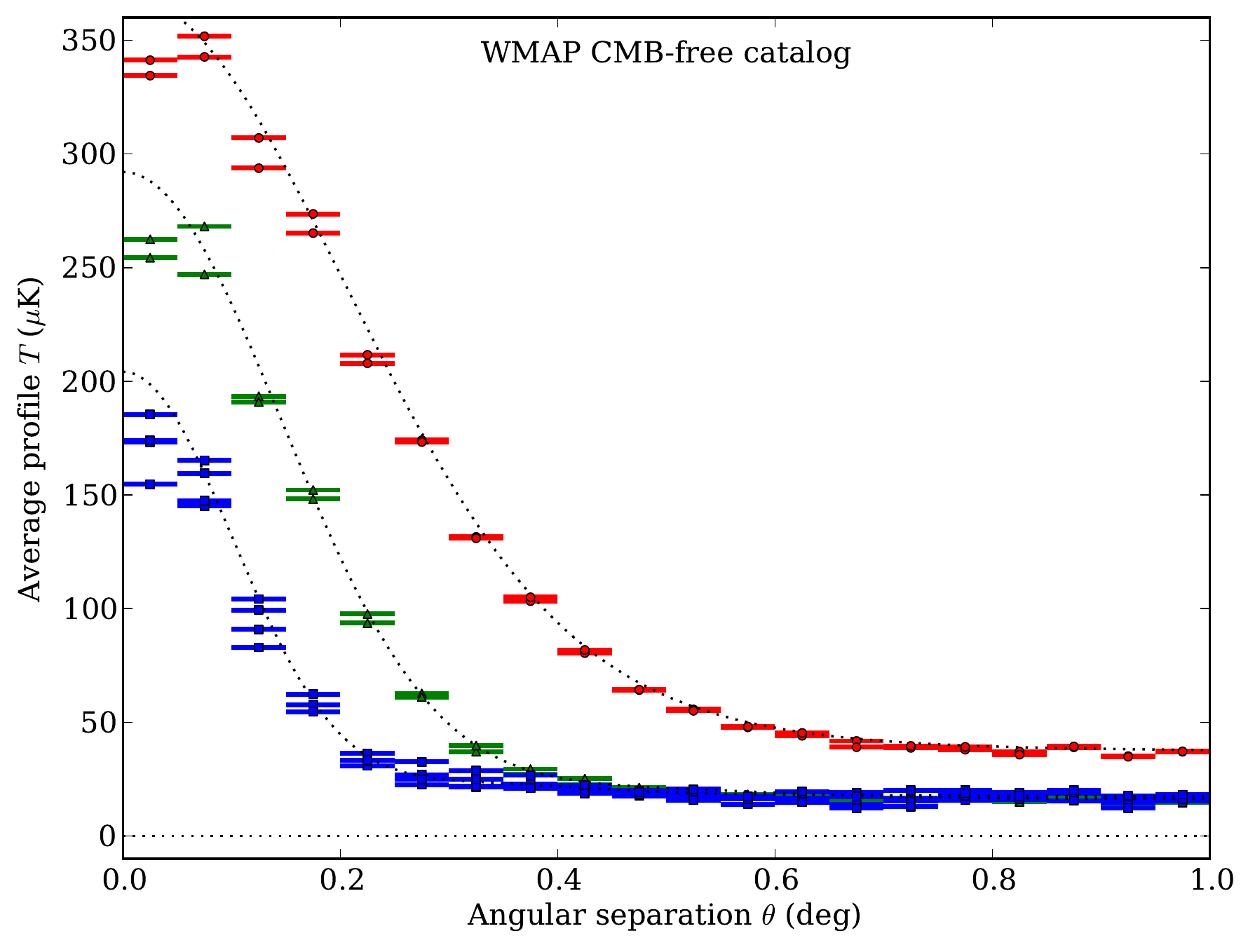}
\hfill
\includegraphics[width=\columnwidth]{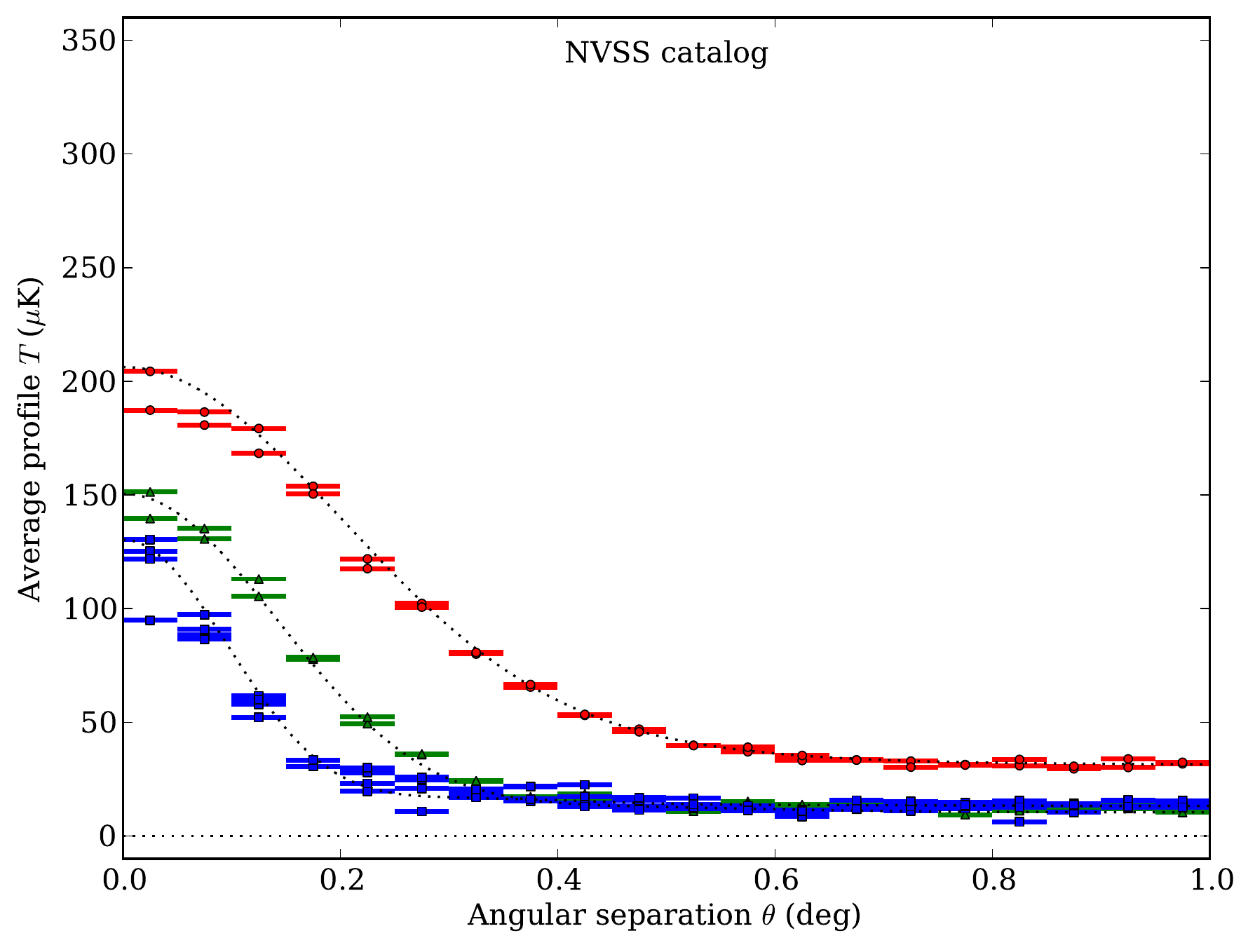}
\caption{Similar to left panel of Figure~\ref{fig:stacked_profiles}, but stacking on the alternative source catalogs that are selected from maps without a significant CMB contribution.  The beam models fit much better here.  Left: WMAP CMB-free catalog.  Right: NVSS catalog.}
\label{fig:all_da_cmbfree}
\end{figure*}
For WMAP catalog sources, we show our stacked profiles from the DA maps in Figure~\ref{fig:stacked_profiles}.  All plots are in thermodynamic temperature and we leave off the error bars to avoid crowding the plot, but show them below.  Consistent with the spectral energy distribution typical of these sources, the average profile is brighter in Q band than in V band, and brighter in V than in W.   Below we quantify that the profiles are significantly wider than the Jupiter-modeled beams.  The heavy tails in the profiles appear similar in V and W at $\theta > 0.4^\circ$, as if the beam profiles are sitting atop a common fluctuation.

We see large scale offsets out to 1 degree, the same as \citet{2010MNRAS.tmpL..93S}.  The offset is $30$--$40$ $\mu$K in Q band and $10$--$20$ $\mu$K in V and W. \citet{2010MNRAS.tmpL..93S} attribute this offset to large scale CMB fluctuations, and fit it at large angular scales.  We disagree with this interpretation.  CMB fluctuations would not cause the substantially larger offset in Q band.  Since the same  sources are stacked in each band, the underlying CMB fluctuations should be the same, up to the effects of beam smoothing.  That Q band is so much higher suggests that galactic foregrounds with a steeply falling spectrum, like synchrotron, might be responsible for this offset.  Indeed the profiles computed in the foreground-reduced maps do not show the large scale offsets.  Here, the Q band profiles too seem to be sitting atop the common fluctuation, joining V and W at $\theta > 0.5^\circ$.

Despite some useful features,  the foreground-reduced maps are  inappropriate for probing the stacked profiles in detail.  The templates for foreground removal also subtract away a portion of each source's flux.  One foreground template is constructed from the K and Ka bands, and when the template is subtracted it creates a depression around each source which is the size of the larger beam scale in those two channels.

The WMAP CMB-free catalog also selects sources that are bright in the WMAP bands;  nearly 80 percent of the sources are the same.  These stacks (Figure~\ref{fig:all_da_cmbfree}) have similar peak temperatures to the WMAP sources at Q-band, and are slightly lower in V and W.   Furthermore, the tails of these profiles are much less heavy than for sources from the standard catalog.

The stacked profiles around NVSS catalog sources \label{fig:all_da_nvss} show that on average the NVSS sources are dimmer than the WMAP catalog sources, and the signal-to-noise is lower.  This is consistent with their selection at lower frequency (at 1.4 GHz), so that more falling spectrum sources and fewer GHz-peaked sources are represented here.  As expected, the same pattern of source peak temperature dropping from Q to V to W holds. 

\subsection{Parametric beam fits}

We fit a two-parameter model to the stacked profiles:
\begin{equation}
m(\theta) = S b(\theta) + m_0
\end{equation}
where $S$ is the amplitude, $b(\theta)$ is the Jupiter-modeled WMAP beam, normalized and smoothed with the pixel window function and Gaussian positional uncertainty, and $m_0$ is an offset to represent large scale foreground or CMB contamination.  Comparisons to the data use the model after binning.
We enforce model beam normalizations so that the 2D integrals under beams are unity,
\begin{equation}
  2\pi \int  d\theta\ \theta\ b(\theta) = 1.
\end{equation}
Thus  $b(\theta)$ has the units of inverse solid angle.  Therefore the amplitude $S$ has units of temperature times solid angle, or equivalently flux density.   For each DA, conversions to flux densities ($dB/dT$) are computed at the effective frequencies given in \citet{2003ApJS..148...39P} for synchrotron-type falling spectrum emission.  

\begin{figure*}
\includegraphics[width=\textwidth]{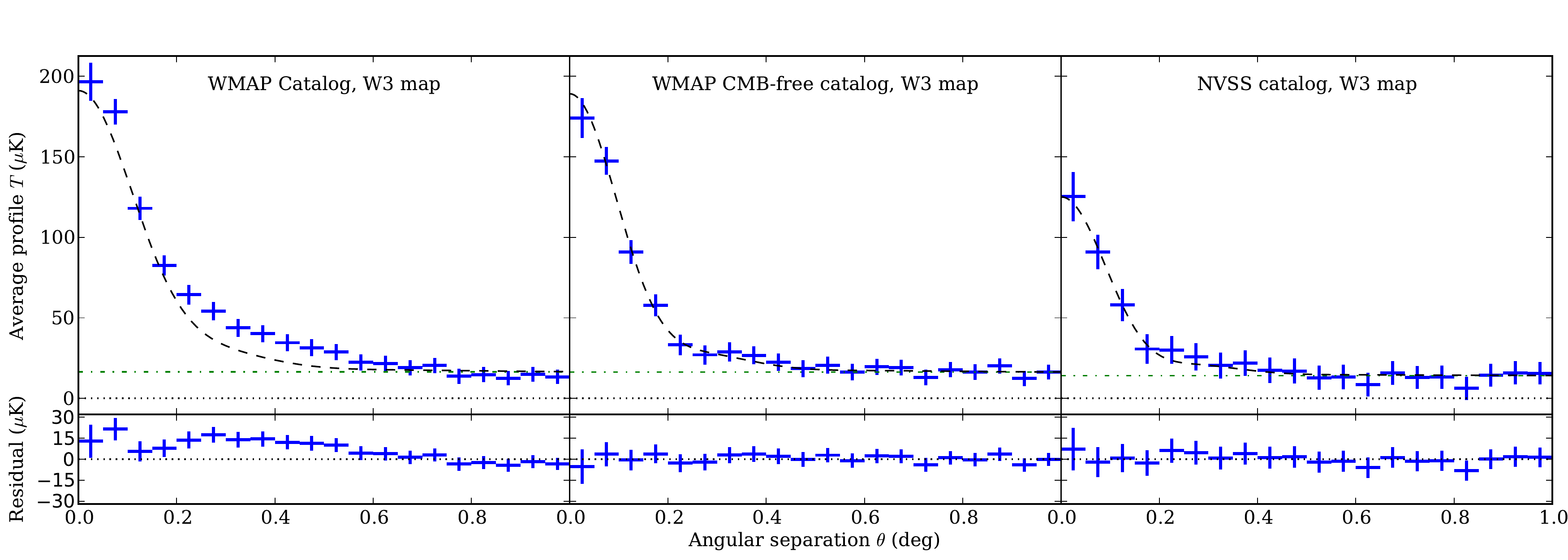}
  \caption{Stacked profile of catalog sources on the W3 WMAP map, fit with the amplitude-offset model (which includes the pixel window function and positional uncertainty).  Plotted errors are from the diagonal of the covariance matrix.   Top left: WMAP standard catalog sources, where the profile shows a bias.  Top middle: WMAP CMB-free catalog sources.  Top right: NVSS catalog sources.   Bottom: binned residuals for each fit.}\label{fig:W3_fits}
\end{figure*}

\begin{figure*}
\includegraphics[width=\textwidth]{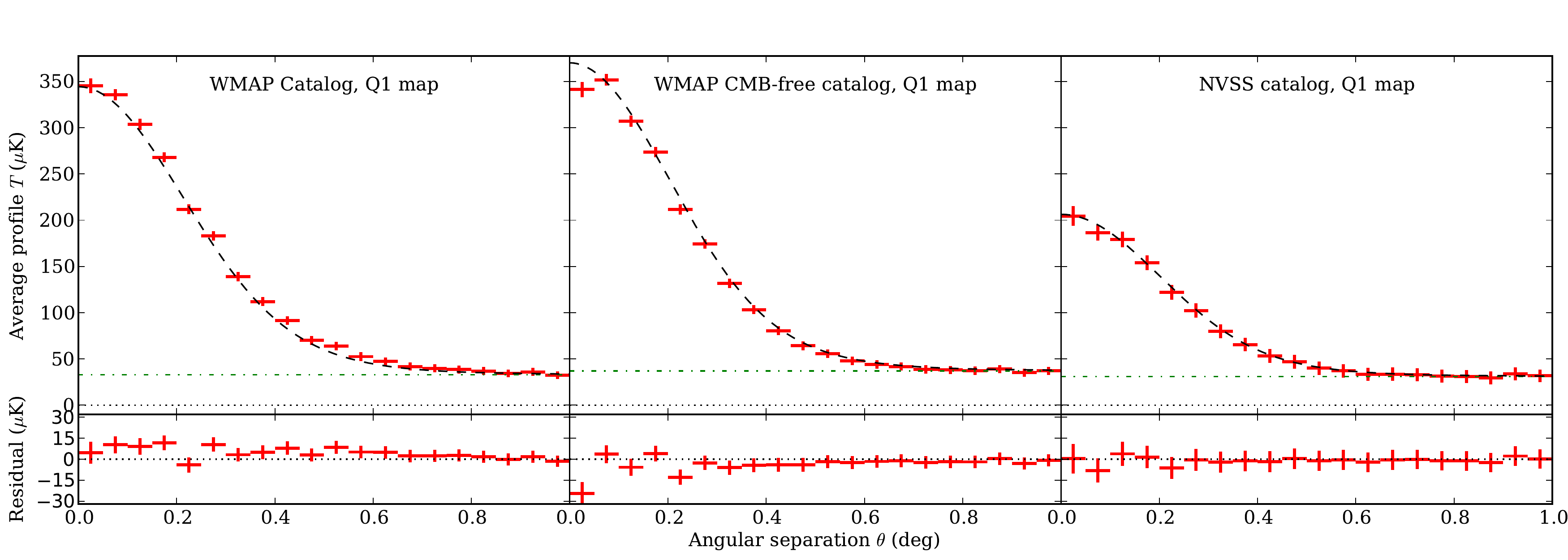}
  \caption{Like Figure~\ref{fig:W3_fits}, stacked profile of catalog sources but on the Q1 WMAP map.  Top left: WMAP standard catalog sources, where the profile shows a bias.  Top middle: WMAP CMB-free catalog sources, where $\chi^2 = 35.41$ for 18 degrees of freedom is particularly high for this catalog.  Top right: NVSS catalog sources.   Bottom: binned residuals for each fit.} \label{fig:Q1_fits}
\end{figure*}

We first examine the best-fit models for the W3 DA; the behavior is similar for the other DAs in the V and W bands.  Figure \ref{fig:W3_fits} shows the best-fit model for this DA and the model's residuals for each of the three catalogs.

The stack on the WMAP standard catalog shows a heavy tail compared to the fit, with large positive residuals.  One seeming peculiarity of this fit is that the fitted model is below the data at most scales, and at least the model's peak could be brought closer to the data by increasing the amplitude.  Such a change nonetheless worsens the $\chi^2$ of the fit, and the strong off-diagonal components in the covariance matrix caused by CMB fluctuations (Figure~\ref{fig:profile_covariance}) are responsible.  Because, unlike the stacked data, the WMAP beam models do not have such heavy tails, the fit will fall below the data at 0.5 degrees for any reasonable amplitude.  Different bins are so strongly correlated that this in turn causes the $\chi^2$ fit to prefer that all bins be below the stacked profile, including those at $\theta < 0.2^{\circ}$.  Boosting the amplitude to raise the model near the peak, which naively would appear to improve the fit, actually worsens $\chi^2$.  Because of the bin-to-bin correlations, in this case the $\chi^2$ fit prefers, in order: (1) all bins low, (2) only some bins low, (3) some bins high and some bins low.  The other DAs show this same behavior stacked on the WMAP standard catalog, except for Q-band fitted to the stack from the foreground-reduced maps where the residuals are negative and caused by the foreground template over-subtraction (and the same reasoning regarding correlated bins applies).

By contrast, the profiles from the CMB-free and NVSS catalogs and  do not show heavy tails.  Remarkably, the shoulder in the W3 beam appears to be recovered in the CMB-free profile.  The NVSS source profiles have lower signal to noise, but are similar.  The other DAs show the same for the CMB-free and NVSS catalogs.  

Thus, sources selected from maps that contain CMB produce stacked profiles with a bias, while sources selected from maps without a significant CMB contribution do not.  These results differ from what \citet{2010MNRAS.tmpL..93S} and \citet{2011arXiv1107.2654W} report for their CMB-free and NVSS stacks.  Below we find this bias is roughly consistent with the expectations of source selection bias due to CMB fluctuations.

This bias for the WMAP standard catalog causes the $\chi^2$ of the fit to be very poor in every DA, seen in Table~\ref{tab:2par_model}, which also shows the fitted parameters and the probability to exceed $\chi^2$ by chance for 18 degrees of freedom (20 bins with two parameters).
  
\begin{table}
  \begin{center} 
    \input{summary_with_errors_and_foreground_poserr.tex}
  \end{center}
  \caption{Parameter values and the goodness-of-fit for a simple amplitude-offset model based on the WMAP beam, for stacked profiles from the three different catalogs.  The probability to exceed $\chi^2$ indicates that the model fits the WMAP catalog profiles poorly, the CMB-free catalog profiles much better (although Q band's probability is low), and the NVSS profiles well. }\label{tab:2par_model} 
\end{table}

The CMB-free catalog probabilities are much more reasonable, except at Q band, where they are a little low.  The probabilities for the NVSS catalog are everywhere reasonable. 

Despite some minor complications which we discuss below, we find no compelling evidence from this modeling that the the Jupiter-based WMAP beams are radically insufficient to explain the source profiles, and we conclude that WMAP's beams for V- and W-bands, which are used for the cosmological analysis, are sound at this level.  Furthermore, the discrepancy with the profiles from the standard WMAP catalog bears the mark of source selection bias due to CMB fluctuations.  We explore this below.

\subsection{CMB-free catalog residuals}

The CMB-free catalog profiles have some features that warrant further discussion.  For example,  for every CMB-free profile, the first bin is low compared to the models for each DA (Figure~\ref{fig:all_da_cmbfree}, left panel).

Figure~\ref{fig:Q1_fits} depicts the profiles and fits for Q1, which is the poorest fit for the CMB-free catalog.  Here the residuals show some interesting patterns.  In both Q-band DAs, the model for the CMB-free catalog exceeds the data slightly at $0.3^\circ < \theta < 0.5^\circ$, leading to negative residuals.  One possibility for this relates to the positional uncertainty.  Brighter sources will have more accurate measured positions, so stacked profiles, weighted toward the bright sources, should have less effective positional uncertainty than the catalog overall (quoted as $2'$ for this case).  In addition, the mix of bright sources in each stack changes from band to band based on the frequency dependence of each source, so even considering a single catalog, the effective positional uncertainty can be different between the bands.

When we allow the positional uncertainty to float in the Q-band fit, the  $\chi^2$ minimum has positional uncertainty less than $2'$, but it is not consistent between Q1, where the fit prefers zero positional uncertainty, and Q2, which prefers $1.9'$, and the probability to exceed $\chi^2$ is still low, never exceeding a couple percent (now for 17 degrees of freedom).  Changing the positional uncertainty from $0'$ to $2.5'$ leads to a 5 percent increase in the inferred source amplitude.  Further letting the model adjust the positional uncertainty in a flux-dependent way makes the model so flexible that it is difficult to draw any conclusions.

The other peculiar feature in Figure~\ref{fig:Q1_fits} for the CMB-free Q1 residuals is the alternating low-high pattern in the first several bins, and the first bin is quite low.  This is visible in the first five bins in each DA except for W1 and W4, where only the first three and two bins (respectively) follow the pattern.  This cannot be ringing in the maps due to the source signal, because the pattern is absent in the NVSS profiles.   Positional uncertainty or systematic mis-centering of the sources do not cause an alternating pattern.

CMB fluctuations, which are common to all the DA maps (up to beam smoothing), are another candidate.  The covariance matrix (Figure~\ref{fig:profile_covariance}) shows that bins 1-3-5 are more strongly correlated to each other (30--35 $\mu K^2$) than they are to bins 2-4 (20--25 $\mu K^2$), so CMB fluctuations could conceivable have an effect shaped like this (with the offset removing the fluctuation common to all bins) and be common across the bands.  However, such an effect should only appear at the couple of $\mu$K level, smaller than what is seen here, and would not be particularly stronger in Q.  

Slight underestimates of the covariance due to positional uncertainty and the finite number of sources,  which would be stronger in Q, are another possibility, but we have not explored it in detail.

\subsection{Flux density and spectral index}

 The source amplitudes are similar in Q band for the WMAP standard and CMB-free catalogs in Table~\ref{tab:2par_model}, but the amplitudes in V and W are substantially less in the CMB-free catalog.  The selection bias causes the sources in V and W to appear much too bright, and this has a significant effect on the measured frequency dependence of sources.

\begin{table}
\begin{tabular}[width=\columnwidth]{lccc}
\hline
Catalog &
$\alpha_{\rm QV}$ &
$\alpha_{\rm VW}$ &
$\alpha_{\rm QW}$
\\  \hline
WMAP ($^*$) &
$-0.11$ $\pm$ $0.03$ &
$-0.07$ $\pm$ $0.04$ &
$-0.09$ $\pm$ $0.02$ 
\\
CMB-free &
$-0.36$ $\pm$ $0.03$ &
$-0.50$ $\pm$ $0.05$ &
$-0.43$ $\pm$ $0.02$ 
\\
NVSS &
$-0.40$ $\pm$ $0.06$ &
$-0.31$ $\pm$ $0.09$ &
$-0.35$ $\pm$ $0.04$  \\
\hline
\end{tabular}
\caption{Frequency scaling for the three catalogs.  The errors represent the error on the catalog averages propagated from Table~\ref{tab:2par_model} and not the much larger intrinsic scatter for sources.\newline
($^*$) The WMAP standard catalog values are biased by the poor fits and shown only for comparison.} \label{tab:alpha}
\end{table}

From the best-fitted source amplitudes, we can examine the scaling for the mean source flux density, writing
\begin{equation}
S_{\rm V} / S_{\rm Q} = (\nu_{\rm V} / \nu_{\rm Q})^{\alpha_{\rm QV}}
\end{equation}
and so forth.  These are displayed in Table~\ref{tab:alpha}.  For the WMAP standard catalog, where we know the flux densities to be biased, we find a mean spectral index $\alpha \sim -0.09$, the same as \citet{2009ApJS..180..283W} found on a source-by-source basis.  However, the spectral index from the CMB-free catalog is much steeper and shows steepening above 61 GHz, from $\alpha_{\rm QV} = -0.36 \pm 0.03 $ to $\alpha_{\rm VW} = -0.50 \pm 0.05$, a difference significant at the $\sim 5\sigma$ level.  The NVSS sources have similarly steeper indices than the WMAP standard catalog sources, but there is no significant change in the spectral index with frequency. 

 These frequency scalings are employed in the estimation of the power spectrum of unresolved point sources, which is necessary to correct the CMB power spectrum.  The source correction effectively changes the tilt of the measured power spectrum.  WMAP's CMB power spectrum results, which are the basis of the cosmological measurement, employ a combination of V- and W-band data only. 

The point source correction uses multifrequency estimators that scale the unresolved sources from Q to the V and W bands.  If the power in Q is held constant, steepening the spectral index according to the CMB-free catalog values reduces the required correction in V by about 20 percent and in W by about 40 percent.  Comparing to \citet{2006ApJ...651L..81H,2008ApJ...688....1H} such changes should raise the value of the scalar index for primordial perturbations ($n_s$) by $\sim 0.01$ or so, which would affect the statistical significance for WMAP's confirmation of the inflationary prediction that $n_s<1$ \citep[see][which gives $n_s = 0.968 \pm 0.012$]{2011ApJS..192...18K}.  However, some caution is appropriate:  the unresolved source contamination is slightly smaller in the more recent WMAP release, making it less susceptible, and this is a substantial enough change to the scaling that a more complete analysis is really required to be quantitative.

\subsection{Selection bias from CMB anisotropy} \label{sec:selectionbias}

The same fluctuations that are responsible for biases in the catalog selection and source counts \citep{1913MNRAS..73..359E} will bias the source profiles.  For the WMAP catalogs, the sources are mostly selected at lower frequency, where they tend to be brightest, and there is a significant mismatch between the beam scales at K-band ($49'$ FHWM) and the Q, V, and W bands ($29'$, $20'$, $13'$ FWHM, respectively)  \citep{2003ApJS..148...39P}.  Although the noise is distinct, the same CMB is seen by each of the DAs (up to beam smoothing).   The profiles in the higher-frequency DAs can therefore be broadened by CMB fluctuations that cannot be distinguished from source flux in the lower frequency maps.

To roughly quantify the bias as a function of source flux density, we performed Monte Carlo simulations, injecting sources of known flux into synthetic realizations of the CMB and detector noise and then finding these sources.  

Following the WMAP source detection procedure \citep{2009ApJS..180..283W}, the maps are weighted by $(N_{\rm obs})^{1/2}$, then filtered in harmonic space by $b_l/(b_l^2w_l^2C^{\rm cmb}_l+N_l)$ to maximize the signal-to-noise of the recovered sources.  The noise spectrum is computed from the pixel noise, pixel area, and the mean over the sky of the inverse number of observations: $N_l = \sigma_0^2 \Omega_{\rm pix} \overline{N_{\rm obs}^{-1}}$.
The pixel noise factors $\sigma_0$ are given by the WMAP data release for each DA.  

\begin{figure}
\includegraphics[width=\columnwidth]{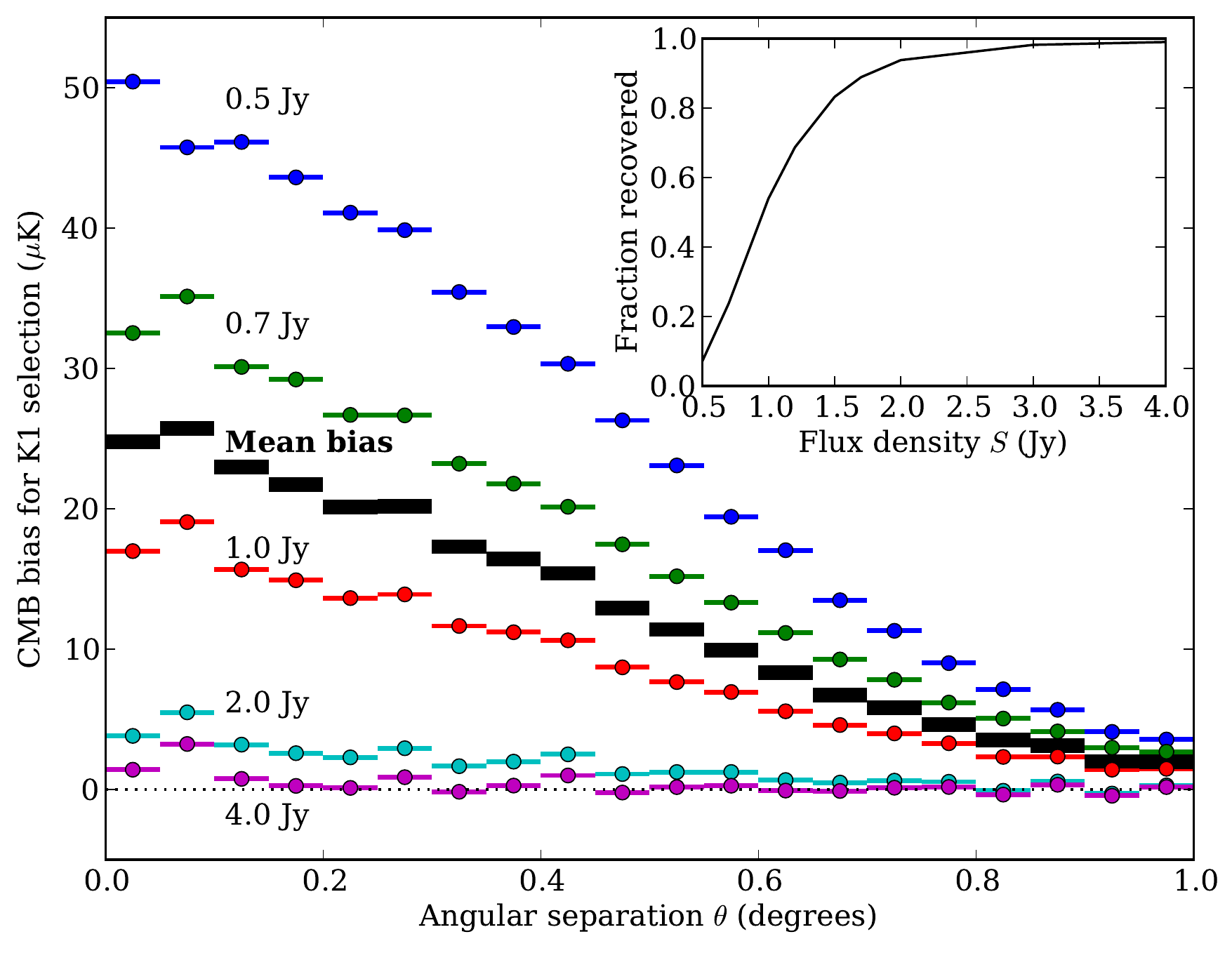}
\caption{Profiles of CMB fluctuations at the location of found sources, for sources with input $S = 0.5$ Jy--$4.0$ Jy in WMAP's K1 band.  This represents a bias in the source profile due to the source selection.  We use the same set of CMB and noise realizations in each case, accounting for some common fluctuations in the curves. Although $\sim 100$ percent of sources with $S> 4.0$ Jy are detected, slight biases remain due to position errors.  The mean profile bias ($P_{\rm bias}(\theta)$) is roughly what we expect for the set of sources in the WMAP catalog.
} \label{fig:selection_bias}
\end{figure}

The synthetic sources that exceed 5$\sigma$ in the filtered map are taken to exceed the catalog threshold.  Our synthetic source selection is slightly simpler than the WMAP selection.  First, we select only in K (where most sources are brightest), instead of the five-band selection used by WMAP.  Second, we use the center of the local maximum pixel as the source position, rather than fitting for the position of each source candidate.   We stack the unfiltered maps around the recovered sources and examine the source profiles.  We show the bias for several input source flux densities in Figure~\ref{fig:selection_bias}.   For a 1.0 Jy source the bias is a little less than 20 $\mu$K at the source position and falls to half-maximum at about 0.5 degrees.  For faint sources, only those that fall on background peaks are recovered, so the bias is large.  Almost all bright sources are found, but slight biases remain because of errors in the source positions, which are shifted to favor background peaks.

After evaluation at several flux densities, we can interpolate to approximate $P_{\rm bias}(\theta,S)$, the bias in the profile as a function of flux density.  Then the expected bias in a catalog can be estimated with 
\begin{equation}
P_{\rm bias}(\theta) = \frac{1}{N_{\rm cat}} \int dS\, \frac{dN_{\rm cat}}{dS} P_{\rm bias}(\theta,S),
\end{equation}
where $dN_{\rm cat}/dS$ gives the distribution of fluxes in the catalog, accounting both for the intrinsic source counts and the selection function, and $N_{\rm cat}$ is the total number of sources.  We estimate this integral by evaluating a sum over sources in the WMAP CMB-free catalog, which is free from CMB contamination and lists noise-bias corrected source fluxes:
\begin{equation}
P_{\rm bias}(\theta) \approx \frac{1}{N_{\rm cat}} \sum_i P_{\rm bias}(\theta,S_i).
\end{equation}
The median K-band flux in this catalog is 0.9 Jy and the minimum de-boosted flux is 0.1 Jy.  The resulting expected bias is about $25$ $\mu$K at the peak and quite broad.  Despite the crudeness of our estimate, which is slightly larger than the residual for the WMAP standard catalog profile in Figure~\ref{fig:W3_fits}, where the fit takes up some of the residual, the bias due to source selection appears to be the probable explanation for the broad profiles found here and in \citet{2010MNRAS.tmpL..93S} for the WMAP standard catalog.

\section{Conclusions} \label{sec:conclusions}
We stacked point sources from three different catalogs on the maps from eight of WMAP's differencing assemblies in the Q, V, and W bands.  For the WMAP standard catalog, we see evidence for residual CMB fluctuations that bias the profiles.  Some complications remain in the profiles for CMB-free catalog sources, but for V and W band the fits are reasonable.  For NVSS sources, the fits are reasonable for all DAs.  Therefore, when sources are selected from data that contain no significant CMB contribution, we find no compelling evidence that the beams for WMAP differ substantially from the Jupiter-based models.  These conclusions directly contradict those of \citet{2010MNRAS.tmpL..93S} and \cite{2011arXiv1107.2654W},  although they do not report the statistical significance of their result.  The reason for the discrepancy, especially for the CMB-free and NVSS catalogs, is not clear.  One possibility relates to the culling of close pairs, where we cut at 2 degrees, while \cite{2011arXiv1107.2654W} cuts NVSS pairs at 1 degree and WMAP CMB-free pairs not at all.  At the same time, they follow the profile out further than we do, beyond two degrees.  Their method for background subtraction also differs from our offset fit.

The biases in the profiles erroneously boost the inferred source flux, especially for the smaller beams, and this affects the source spectral index.  Using fluxes from the CMB-free catalog, the spectral indices are significantly steeper, and show spectral steepening at high frequency.  This in turn affects the point source subtraction for the power spectrum.

The issues we confronted in this work (pixel and binning effects, positional uncertainty, foreground subtraction, selection bias) are common to all microwave experiments that construct point source catalogs.  Recent catalogs from ACT \citep{2011ApJ...731..100M} and SPT \citep{2010ApJ...719..763V} are much less susceptible to the CMB fluctuations because the power in the CMB falls so rapidly on arcminute angular scales. These catalogs take care to de-boost their measured flux densities for the biases due to noise and other sources.
However, at these frequencies and angular scales, the background of dusty galaxies can play a role similar to that the CMB plays here.  Subsequent analysis of these catalogs could run into similar biases, for example, when stacking 220 GHz maps on 150 GHz source positions or vice versa, or stacking ACT 220 GHz sources in the SPT data where the beam is smaller.

Planck data \citep{2011arXiv1101.2022P}, with larger beams ($5'$--$30'$), is more susceptible to CMB fluctuations, so the same cautions apply as for WMAP, both for the recent Early Source Catalog \citep{2011arXiv1101.2041P} and for the final band-merged catalog.
 
\section*{Acknowledgments}
We thank Xi Chen, Reijo Keskitalo, Mike Nolta, and Ned Wright for useful comments and discussion.
Some of the results in this paper have been derived using the HEALPix package \citep{2005ApJ...622..759G} and Healpy. This work was partially supported by a Space Education and Training Program award from the Florida Space Grant Consortium.
KWS acknowledges the support of the Beyond the Book program from the University of Miami College of Arts and Sciences.
KMH receives support from NASA-JPL subcontract 1363745.

\appendix

\input{appendix}

\bibliographystyle{mn}
\bibliography{notes.bib}

\end{document}

%% file: summary_with_errors_and_foreground_poserr.tex
\begin{tabular*}{\columnwidth}{c c c c c}
\multicolumn{5}{c}{WMAP catalog}\\
\hline
DA & Amp. (Jy) & Offset ($\mu$K) & $\chi^2_{\nu=18}$ &  $P( > \chi^2)$ \\ [0.5ex]
\hline
Q1 & 1.51 $\pm$ 0.01 & 32.80 $\pm$ 4.07 & 57.08 & $6.01\times 10^{-6}$ \\
Q2 & 1.51 $\pm$ 0.01 & 33.45 $\pm$ 4.17 & 44.24 & $5.34\times 10^{-4}$ \\
V1 & 1.46 $\pm$ 0.02 & 12.49 $\pm$ 4.27 & 49.45 & $9.12\times 10^{-5}$ \\
V2 & 1.43 $\pm$ 0.02 & 13.97 $\pm$ 4.14 & 67.23 & $1.32\times 10^{-7}$ \\
W1 & 1.38 $\pm$ 0.04 & 16.57 $\pm$ 4.33 & 56.09 & $8.63\times 10^{-6}$ \\
W2 & 1.43 $\pm$ 0.05 & 16.92 $\pm$ 4.27 & 69.83 & $4.83\times 10^{-8}$ \\
W3 & 1.41 $\pm$ 0.05 & 16.39 $\pm$ 4.19 & 58.66 & $3.36\times 10^{-6}$ \\
W4 & 1.38 $\pm$ 0.04 & 16.54 $\pm$ 4.21 & 47.35 & $1.88\times 10^{-4}$ \\
\hline
\\
\multicolumn{5}{c}{WMAP CMB-free catalog}\\
\hline
DA & Amp. (Jy) & Offset ($\mu$K) & $\chi^2_{\nu=18}$ &  $P( > \chi^2)$ \\ [0.5ex]
\hline
Q1 & 1.50 $\pm$ 0.01 & 36.97 $\pm$ 4.28 & 35.41 & $8.39\times 10^{-3}$ \\
Q2 & 1.49 $\pm$ 0.01 & 36.58 $\pm$ 4.27 & 30.34 & $0.034$ \\
V1 & 1.31 $\pm$ 0.02 & 15.97 $\pm$ 4.26 & 25.52 & $0.111$ \\
V2 & 1.27 $\pm$ 0.02 & 15.56 $\pm$ 4.39 & 19.22 & $0.378$ \\
W1 & 1.04 $\pm$ 0.03 & 17.32 $\pm$ 4.27 & 16.86 & $0.533$ \\
W2 & 1.05 $\pm$ 0.04 & 16.85 $\pm$ 4.40 & 24.60 & $0.136$ \\
W3 & 1.03 $\pm$ 0.04 & 16.25 $\pm$ 4.35 & 17.16 & $0.512$ \\
W4 & 1.05 $\pm$ 0.04 & 17.23 $\pm$ 4.32 & 8.89 & $0.962$ \\
\hline
\\
\multicolumn{5}{c}{NVSS catalog}\\
\hline
DA & Amp. (Jy) & Offset ($\mu$K) & $\chi^2_{\nu=18}$ &  $P( > \chi^2)$ \\ [0.5ex]
\hline
Q1 & 0.77 $\pm$ 0.01 & 31.04 $\pm$ 6.72 & 17.05 & $0.520$ \\
Q2 & 0.74 $\pm$ 0.01 & 28.86 $\pm$ 6.72 & 24.59 & $0.137$ \\
V1 & 0.63 $\pm$ 0.02 & 10.27 $\pm$ 6.57 & 14.40 & $0.703$ \\
V2 & 0.66 $\pm$ 0.02 & 12.14 $\pm$ 6.73 & 12.43 & $0.825$ \\
W1 & 0.58 $\pm$ 0.03 & 13.17 $\pm$ 6.70 & 12.02 & $0.846$ \\
W2 & 0.54 $\pm$ 0.04 & 12.81 $\pm$ 6.57 & 15.92 & $0.598$ \\
W3 & 0.59 $\pm$ 0.04 & 14.04 $\pm$ 6.77 & 24.41 & $0.142$ \\
W4 & 0.54 $\pm$ 0.04 & 13.81 $\pm$ 6.71 & 9.25 & $0.954$ \\
\hline
\\
\end{tabular*}

%% file: appendix.tex
\section{Covariance for stacked profiles}
\label{sec:covariance_app}
 The stacking procedure is equivalent to the linear operation:
\begin{equation}
  P_i = \frac{\sum_q S_{iq} M_q}{\sum_q S_{iq}} \label{eqn:stack}
\end{equation}
where $P_i$ is the stacked profile in angular bin $i$, $M_q$ is the $q$th pixel's value in the map, and a stacking operator $S_{iq}$ is
\begin{equation}
S_{iq} = \left\{ 
\begin{array}{ll} 
1 & \mbox{if pixel $q$ is in annulus $i$ about a source,} \\ 
0 & \mbox{otherwise.}
\end{array}
\right.
\end{equation}
The catalog point source positions are implicit in the $S$ matrix, thus the contraction of the map with the stacking operator separately accumulates the contribution for each profile bin.

If the map $M$ is zero mean, as for CMB and detector noise, then the covariance in of profile bins is a simple function of the covariance of maps.
\begin{equation}
  {\rm Cov}(P_i,P_j)
  = \frac{\sum_{qr} S_{iq} S_{jr} \langle M_q M_r \rangle}{ \left( \sum_q S_{iq}\right) \left( \sum_r S_{jr} \right)} \label{eq:formalcov}
\end{equation}

For a map with only white detector noise,  $n_q$, we have $\langle n_q n_r \rangle =  \delta_{qr} \sigma_0^2/N_{{\rm obs},q}$.  Therefore, 
\begin{eqnarray}
{\rm Cov}(P_i,P_j) &=& \sigma^2_0 \frac{\sum_{qr} S_{iq} S_{jr} \delta_{qr} N^{-1}_{{\rm obs},q}  }{ \left( \sum_q S_{iq}\right) \left( \sum_r S_{jr} \right)}  \nonumber \\
&=& \sigma^2_0 \frac{\sum_{q} S_{iq} S_{jq} N^{-1}_{{\rm obs},q}  }{ \left( \sum_q S_{iq}\right) \left( \sum_r S_{jr} \right)}
\end{eqnarray}
If additionally we require that the sources are well-separated, that is, that the annuli of different sources never overlap (and the annuli around single sources naturally never overlap), then the bins are uncorrelated and the covariance simplifies:
\begin{equation}
{\rm Cov}(P_i,P_j) = \delta_{ij} \sigma^2_0 \frac{\sum_{q} S_{iq} S_{iq} N^{-1}_{{\rm obs},q}  }{ \left( \sum_q S_{iq}\right) \left( \sum_r S_{ir} \right)}
\end{equation}
Since $ S_{iq}$ has value zero or one, then $ S_{iq}  S_{iq} =  S_{iq}$, so \begin{equation}
{\rm Cov}(P_i,P_j) = \delta_{ij} \sigma^2_0 \frac{\sum_{q} S_{iq} N^{-1}_{{\rm obs},q}  }{ \left( \sum_q S_{iq}\right)^2}
\end{equation}

\begin{equation}
C_{ij} \equiv {\rm Cov}(P_i,P_j) = \delta_{ij}  \sigma_0^2 \frac{\sum_q S_{iq} N^{-1}_{{\rm obs},q}}{\left(\sum_q S_{iq} \right)^2} \hfill \mbox{(noise only)} \quad
\end{equation}
where $\sigma_0^2  N^{-1}_{{\rm obs},q}$ is the noise variance in pixel $q$, provided by the WMAP team for each differencing assembly's map.  Noting the similarity to the stacking operation (Eq.~\ref{eqn:stack}), we compute this quantity using a slight modification to the stacking code, but applied to the $N_{\rm obs}^{-1}$ map with a different normalization, and verify with Monte Carlo noise simulations.

For the CMB, the term in angle braces in (\ref{eq:formalcov}) is simply the angular correlation function, obtained from the theory angular power spectrum $C_l$ by a sum over Legendre polynomials:
\begin{equation}
 \langle M_q M_r \rangle = \omega(\theta_{qr}) = \frac{1}{4 \pi} \sum_l  \left( 2l+1 \right) b^2_l w^2_{l} C_l  P_l(\cos \theta_{qr}).
\end{equation}
The beam window function for the map is represented by $b^2_l$ and the pixel window function by $w^2_{l}$.  In practice, however, it is computationally inconvenient to keep track of all possible pixel separations ($\theta_{qr}$), and more convenient to compute this portion of the covariance using a Monte Carlo of CMB realizations.